\documentclass[reprint,prc,showpacs,preprintnumbers,amsmath,amssymb]{revtex4-1}
\usepackage{graphicx}
\usepackage{dcolumn}
\usepackage{amsmath}
\usepackage{times}
\usepackage{mathtools}

\begin{document}

\title{Approximations of potentials through the truncation of their inverses }

\author{N.~C.~Brown}
\author{S.~E.~Grefe }
\author{Z.~Papp}
\affiliation{ Department of Physics and Astronomy,
California State University Long Beach, Long Beach, California, USA }

\date{\today}

\begin{abstract}
The inverse of an $\infty \times \infty$ symmetric band matrix can be constructed in terms of a matrix continued fraction. For Hamiltonians with Coulomb plus polynomial potentials, this results in an exact and analytic Green's operator which, even in finite-dimensional representation, exhibits the exact spectrum. 
In this work we propose a finite dimensional representation for the potential operator such that it retains some information about the whole Hilbert-space representation. The potential should be represented in a larger basis, 
then the matrix should be inverted, then truncated to the desired size, and finally inverted again. 
This procedure results in a superb low-rank representation of the potential 
operator. The method is illustrated with a typical nucleon-nucleon potential.

\end{abstract}

\pacs{ 21.45.-v, 03.65.Ge, 03.65.Nk, 03.65.Aa, 02.30.Rz, 02.30.Mv  }

\maketitle

\section{Introduction}
\label{intro}
A quantum mechanical system is completely described by the Hilbert space and by the Hamiltonian $H$. 
However, in practical calculations, the infinite dimensional Hilbert space is often truncated
and the calculations are performed on a finite subset of the basis. This truncation is variational and
 the exact results are reached as the truncated basis approaches the complete basis set.

Green's operators are defined by $G(z)=(z-H)^{-1}$, and they can also be used to describe the 
quantum system. The poles of the Green's operator are the eigenvalues of the Hamiltonian and the 
residue are the projection onto the subspace spanned by the eigenfunctions. However, if at some 
energy, the Green's operator is singular, then it is
singular in any representation and thus also singular in a finite subspace representation. The poles of the 
Green's operator are insensitive to the truncation of the Hilbert space. 

In general, it is much harder to work with the Green's operator than with the Hamiltonian. The Hamiltonian can be represented by differential operators, while the Green's operator, the inverse of the Hamiltonian, is an integral operator. 
If the Hamiltonian is represented in a countable infinite basis by an $\infty \times \infty$ matrix, the Green's operator is the inverse of an $\infty \times \infty$ matrix. 
Working with $\infty \times \infty$ matrices is not very encouraging, but they can be inverted in special cases. If the matrix $J=z-H$ is of Jacobi type, 
i.e. it is an infinite symmetric tridiagonal matrix, then an $N\times N$ representation of the Green's operator,  $G^{N}$, can be constructed from the 
$N\times N$ representation of $J$ plus a continued fraction.
 In fact $(G_{N})^{-1}$ is 
almost identical to $J_{N}$,
the only difference being a continued fraction is added  to the $N\times N$ term. This is the way
the Coulomb Green's operator has been determined in the Coulomb-Sturmian  basis representation. 
\cite{Konya:1997JMP,Konya:1999by,PRADemir2006}. 

The approach has been extended to infinite symmetric band matrices. An infinite symmetric band matrix can be considered as a Jacobi matrix of block matrices. 
Therefore 
\begin{equation}\label{gm1}
(G^{N})^{-1}_{i,j} = J_{i,j}^{N} - \delta_{i,N}\delta_{j,N} J_{N,N+1} C_{N+1} J_{N+1,N},
\end{equation}
where $J_{i,j}$ are block matrices and $C_{N+1}$ is a matrix continued fraction.
Using this method,  the Green's operators for Hamiltonians
containing kinetic energy, Coulomb and polynomial potentials have been evaluated in Ref.\  \cite{kelbert2007green}. 

 At eigenvalue energies the determinant of the $G_{N}(E)$ is singular and the 
determinant of $(G_{N}(E))^{-1}$ vanishes. So, we can consider $ (G_{N}(E))^{-1}$ as an improved Hamiltonian, which irrespective of $N$, provides the correct eigenvalues. This way we accomplished a kind
of ``packing'' of the $\infty \times \infty$ matrix into an $N \times N$ matrix, where $N$ is not necessarily big.

A general Hamiltonian,  besides the Coulomb and and some polynomial potential, may contain a short-range potential as well.
 A general short-range potential in a discrete basis representation is certainly not tridiagonal, not even 
 block-tridiagonal. But for any reasonable potential, the matrix representation looks like a ridge: the 
 matrix elements are much bigger if $n$ and $n'$ are close and become negligible otherwise, just  
 like a band matrix. Therefore, the technique developed before may be applicable for finding a faithful matrix representation of a general short-range potential. 
 
 The aim of this paper is find a low-$N$ representation of the short-range potential such 
that it carries the information of the whole Hilbert-space representation. 
We accomplish our goal through the inverse of the potential operator. 

We present our results in the Coulomb-Sturmian basis representation, but we believe that the results are valid for 
any discrete basis provided the matrix exhibits a ridge-like structure.
The method of approximating the potential on Coulomb-Sturmian basis has quite a long history. It has successfully been applied to various problems, including the solution of the Faddeev equations with a Coulomb \cite{PhysRevC.54.50,PhysRevC.55.1080} and confining \cite{PhysRevC.62.044004} potential. In Sec.\ \ref{sec2}, we outline the technique for solving the Lippmann-Schwinger equation.  Then in Sec.\ \ref{sec3} we introduce the physical system and outline previous schemes for approximating the potential.  Then in Sec.\ \ref{sec4} we apply the 
inverse matrix idea to matrices with ridge-like structure. Finally we summarize our findings in Sec.\ \ref{summary}.

\section{Solution of the Lippmann-Schwinger equations}
\label{sec2}

We consider a Hamiltonian with a Coulomb $v^{C}$ plus short-range $v^{(s)}$ potential.
The bound states are the solutions of the homogeneous 
\begin{equation}\label{lsb}
|\psi \rangle = g_{l}^{C}(E) v_{l}^{(s)}|\psi \rangle,
\end{equation}
while the scattering states are the solutions of the inhomogeneous
\begin{equation}\label{lssc}
|\psi^{(\pm)} \rangle = |\phi_{l}^{C(\pm)}\rangle + g_{l}^{C}(E\pm i\epsilon) v_{l}^{(s)}|\psi^{(\pm)} \rangle
\end{equation}
Lippmann-Schwinger equations. Here $E$ is the energy, $l$ is the angular momentum, 
$g_{l}^{C}(E)=(E-h_{l}^{0}-v^{C})^{-1}$ is the 
Coulomb Green's operator, $h_{l}^{0}$ is the kinetic energy and $\phi_{l}^{C}$ 
is the Coulomb scattering state.
The scattering state $\psi^{(\pm)}$ is related to the scattering amplitude by
\begin{equation}\label{al}
a_{l}= \langle \phi_{l}^{C(-)} | v_{l}^{(s)} |\psi^{(+)} \rangle = \frac{\exp(i(2\eta_{l}+\delta_{l}))}{k} \sin \delta_{l}~,
\end{equation}
where $k$ is the wave number, $\eta_{l}$ is the Coulomb phase shift  and $\delta_{l}$ is the Coulomb-modified nuclear phase shift.

The Coulomb-Sturmian basis, in angular momentum $l$, is defined by
\begin{equation}
\langle r | n l \rangle = \frac{\sqrt{n!}}{\sqrt{(n+2l+1)!}} \exp(-br) (2br)^{l+1} L_{n}^{2l+1}(2br)
\end{equation}
and
\begin{equation}
\langle p | n l \rangle =  \frac{\sqrt{2} \sqrt{n!}(n+l+1) l! (4bp)^{l+1}}{\sqrt{\pi}   \sqrt{(n+2l+1)! } (p^{2} +b^{2})^{l+2}} G_{n}^{l+1}\left (\frac{p^{2}-b^{2}}{p^{2}+b^{2}}\right),
\end{equation}
in configuration and momentum space, respectively. Here $L$ and $G$ are the Laguerre and Gegenbauer polynomials, respectively, and $b$ is a parameter. Together with 
\begin{equation}
\langle r | \widetilde{nl} \rangle = \langle r |  {nl} \rangle /r
\end{equation}
and 
\begin{equation}
\langle p | \widetilde{nl} \rangle = \langle p |  {nl} \rangle \frac{p^{2}+b^{2}}{2 b (n+l+1)}
\end{equation}
these functions are orthonormal
\begin{equation}
\langle nl | \widetilde{n' l} \rangle = \delta_{n n'}
\end{equation}
and form a complete set
\begin{equation}
\lim_{N\to\infty}\sum_{n=0}^{N} | nl \rangle \langle \widetilde{n' l} | = 1~.
\end{equation}

The finite dimensional representation of the short-range potential is given by
\begin{equation}\label{vapprox}
v_{l}^{(s)} \approx v^{N,N} = \sum_{n n'}^{N} | \widetilde{nl}\rangle  \underline{v}^{(s) N,N}_{l,n n'}  \langle \widetilde{n' l} |~,
\end{equation}
where $ \underline{v}^{(s)N,N}_{l,n n'} = \langle{nl} | v_{l}^{(s)} | n' l \rangle$. 
Now the Lippmann-Schwinger equations (\ref{lsb}) and (\ref{lssc}) become matrix
equations 
\begin{equation}\label{lsbm}
\underline{\psi}   = \underline{g}_{l}^{C} \underline{v}_{l}^{(s)} \underline{\psi}
\end{equation}
and
\begin{equation}\label{lsscm}
\underline{\psi}  = \underline{\phi}_{l}^{C} + \underline{g}_{l}^{C} \underline{v}_{l}^{(s)} \underline{\psi},
\end{equation}
respectively, where the  matrices and vectors are underlined.
Some rearrangement gives
\begin{equation} \label{homae}
( ( \underline{g}_{l}^{C})^{-1}   - \underline{v}_{l}^{(s)} )  \underline{\psi} =  0
\end{equation}
and 
\begin{equation}\label{inhome}
( ( \underline{g}_{l}^{C})^{-1}   - \underline{v}_{l}^{(s)} )  \underline{\psi} =  (\underline{g}_{l}^{C})^{-1} \underline{\phi}_{l}^{C}~,
\end{equation}
i.e.\ the homogeneous Lippmann-Schwinger equation becomes a homogeneous algebraic equation and the 
inhomogeneous Lippmann-Schwinger equation becomes an inhomogeneous algebraic equation. The homogeneous algebraic equation is solvable if the determinant is zero
\begin{equation}
|( \underline{g}_{l}^{C})^{-1}(E)   - \underline{v}_{l}^{(s)} | =0.
\end{equation}
This condition provides the eigenvalues and the solution of (\ref{homae}) provides the eigenvectors. The solution of the inhomogeneous algebraic equation gives the scattering state $\underline{\psi}$, which, with the help of Eq.\ (\ref{al}), can provide us with the phase shift.

The matrix $(\underline{g}_{l}^{C})^{-1}$ can be calculated by using (\ref{gm1})  
\begin{equation}
(\underline{g}_{l}^{C})^{-1} = \underline{J}^{C} - \delta_{i,N}\delta_{j,N} J^{2}_{N,N+1} C_{N+1}~.
\end{equation}
Here $J(E)=(E-h_{l}^{0}-v^{C})$ and $v^{C}=Z/r$.  The matrix $\underline{J}^{C}$ is symmetric tridiagonal, and the nonzero elements  are given by
\begin{equation}
\underline{J}_{i,i}^{C} = 2(i+l+1) \frac{\hbar^{2}(k^{2}-b^{2})}{4\mu b}-Z
\end{equation}
and 
\begin{equation}
\underline{J}_{i,i+1}^{C} = -\sqrt{(i+1)(i+2l+2)} \frac{\hbar^{2}(k^{2}+b^{2})}{4\mu b}~,
\end{equation}
where $\mu$ is the reduced mass and $k=\sqrt{2\mu/\hbar^{2} \:E}$.
In this particular case the continued fraction can be summed up to a ratio of hypergeometric functions
\begin{eqnarray}
C_{N+1} &&=  -\frac{4m/\hbar^{2}\:b }{(b-ik)^{2}(N+l+2+i\gamma) }  \\
&& \times \frac{_{2}F_{1}\left(-l+i\gamma,N+2;N+l+3+i\gamma;\left(  \frac{b+ik}{b-ik} \right)^{2} \right)  }
{ _{2}F_{1}\left(-l+i\gamma,N+1;N+l+2+i\gamma;\left(  \frac{b+ik}{b-ik} \right)^{2} \right)  }~, \nonumber
\end{eqnarray}
where $\gamma=Z\mu/(\hbar^{2}k)$  \cite{PRADemir2006}.
The analytic evaluation of   $\underline{\phi}_{l}^{C}$ has been presented  
before  in Ref.\  \cite{PhysRevC.38.2457}.

\begin{figure}
\centering
\includegraphics[width=8.5cm]{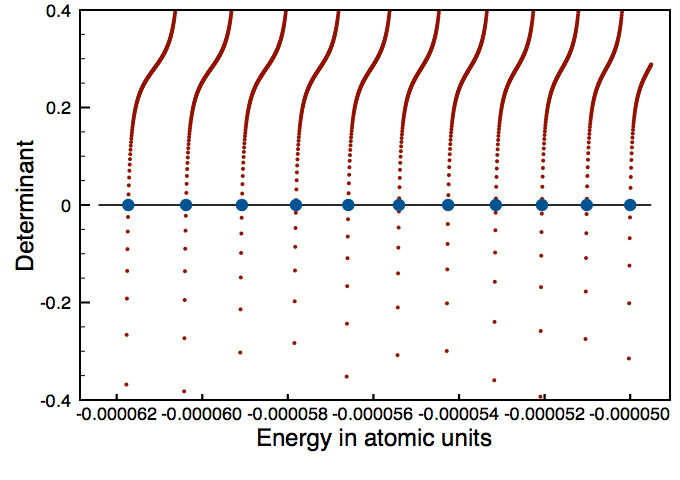}
\caption{The zeros of $\underline{g}_{0}^{C}(E)$ for a hydrogen system in atomic units in the energy
range $E_{90}-E_{100}$. The large dots represent the exact eigenvalues.}
\label{fighyd}
\end{figure}

This representation of $(\underline{g}_{l}^{C})^{-1}$ is exact and analytic. Even a very low-rank matrix gives an account
for the complete spectrum of the Coulomb Hamiltonian. Fig.\ \ref{fighyd} shows the  determinant of the $3 \times 3$ 
$(\underline{g}_{l}^{C})^{-1}$ matrix for Coulomb Hamiltonian with $l=0$, $Z=-1$, $\mu=1$ and $\hbar=1$. The exact eigenvalues 
are $E_{n}=-1/(2 n^{2})$. The figure shows the energy range corresponding to the $E_{90}-E_{100}$ interval. We can
see that the numerical zeros are at the exact locations even in this extreme case.

In this approach the only approximation is the finite-basis representation of the potential, since the evaluation
of $\underline{g}_{l}^{C}$ and $\underline{\phi}_{l}^{C}$ is exact and analytic. Therefore both the bound and scattering state 
wave function $\psi$ and $\psi^{\pm}$
possess the exact Coulomb-like asymptotic behavior \cite{PhysRevA.46.4437}.

Finite-rank potentials have a long history in physics (see eg.\ Ref.\ \cite{adhikari1991dynamical}). Various schemes have been proposed. 
Most of them use some form factors which allow for an easy and exact 
evaluation of the matrix elements of the Green's operator.  The use of Coulomb-Sturmian functions offers several advantages. 
Since they form a basis, the convergence of the approximation is guaranteed. More importantly, 
 it works with Coulomb-like potentials, unlike the majority of approaches.

\section{The example problem}
\label{sec3}

To illustrate the method we consider a typical nucleon-nucleon
 potential, the Malflet-Tjon potential.
This potential has a strong repulsive core and an attractive tail, like most of the potentials in physics.
The Malflet-Tjon potential is given by
\begin{equation}
v^{s}=  v_{1} \exp( - \beta_{1} r )/r + v_{2} \exp( - \beta_{2} r )/r 
\end{equation}
with $v_{1}=1438.720\: \mbox{MeV}$, $\beta_{1}= 3.11\: \mbox{fm}^{-1}$, 
$v_{2}= -626.885 \:\mbox{MeV}$, $\beta_{2}= 1.55\: \mbox{fm}^{-1}$. 
The other parameters in the model are charge parameter $Z=e^{2}=1.44 \:\mbox{MeV fm}$, 
$\hbar^{2}/m = 41.47\: \mbox{MeV / amu}$ and nucleon reduced mass $\mu=1/2\: \mbox{amu}$.
We used $b=3\: \mbox{fm}^{-1}$, which is around the optimum. We note that the rate of convergence is rather insensitive
to the choice of $b$ within a rather broad interval.

\begin{figure}
\centering
\includegraphics[width=8.5cm]{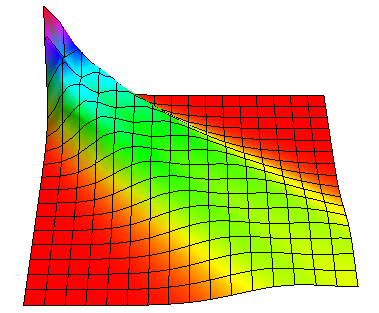}
\caption{$\underline{v}^{20,20}$ Coulomb-Sturmian matrix elements of the Malflet-Tjon potential.}
\label{step0}
\end{figure}

Figure \ref{step0} shows the $\underline{v}^{20,20}$ matrix. We can see that the matrix  representation 
exhibits a ridge-like structure and the dominant matrix elements decrease only very slowly. So, if we
truncate the basis to this size, we chop down the tail of the matrix and we neglect terms which are not 
small at all. Consequently, this representation results in a slow convergence.

\begin{figure}
\centering
\includegraphics[width=8.5cm]{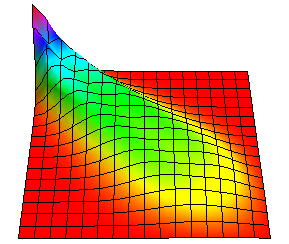}
\caption{Potential matrix $\underline{\tilde{v}}^{20}$. The matrix of Fig.\ \ref{step0} has been modified by the $\sigma$ factors of Eq.\ (\ref{sig}).}
\label{sigma}
\end{figure}

We have to note here that there had been approaches before to improve the situation. Inspired by 
Lanczos filtering, it has been proposed to multiply the potential matrix by some function which suppress the higher elements \cite{revai,gyarmati1979rigorous}
\begin{equation}
\underline{\tilde{v}}^{N}_{i,j} =  \sigma_{i}^{N} \underline{v}^{N,N} \sigma_{j}^{N},
\end{equation}
where 
\begin{equation}\label{sig}
\sigma_{i}^{N} = \frac{ 1-\exp(-[\alpha(i-N-1)/(N+1)]^{2}) }{1-\exp(-\alpha^{2})}
\end{equation}
with $\alpha\sim 6$. This approach results in a transformed matrix shown in Fig.\ \ref{sigma}.

\begin{figure}
\centering
\includegraphics[width=8.5cm]{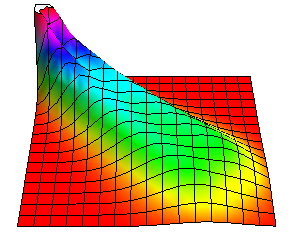}
\caption{Potential matrix $\underline{\hat{v}}^{20}$ from double-basis representation of Eq.\ (\ref{2b}).}
\label{2bases}
\end{figure}

In the other approach two Hilbert-space bases has been adopted 
\cite{PhysRevC.36.1275,PhysRevC.63.057001}
\begin{equation}\label{2b}
\underline{\hat{v}}^{N} = \underline{O} \: \underline{v}^{N,N} \: \underline{O}',
\end{equation}
where $\underline{v}^{N,N}$ has been calculated with basis parameter $b_{1}$ and 
$\underline{O}= (\langle \widetilde{ nl;b_{1}}| n'l;b_{2} \rangle)^{-1}$. 
The  
a potential matrix $\underline{\hat{v}}$ with $b_{1}=2.5\:\mbox{fm}^{-1}$ and $b_{2}=3.5\:\mbox{fm}^{-1}$ is shown in Fig.\ \ref{2bases}. It is interesting to note that this approach also utilizes the inverse of the potential operator \cite{PhysRevC.36.1275}. Our approach is however different, as we are going to see below.

We can see that both methods basically  suppress the higher index elements of the matrix. 
Now the transition to the neglected terms is smooth. We found that among these two methods, the one with
 two bases gives a faster convergence \cite{PhysRevC.63.057001}.

\section{ Approximation through the inverse}
\label{sec4}

We saw before that we could achieve a good approximation of the Hamiltonian by considering 
it on an infinite
basis representation and then by rolling up the tail of the band matrix into a matrix continued fraction. 
Here we try a similar
procedure with the potential operator. First we calculate the Coulomb-Sturmian matrix elements of ${v}^{s}$ on a basis of $N'$ size, invert the matrix, then 
truncate it to $N\le N'$, and finally invert the matrix again. We denote the resulting matrix by 
$\underline{v}^{N,N'}$.

\begin{figure}
\centering
\includegraphics[width=8.5cm]{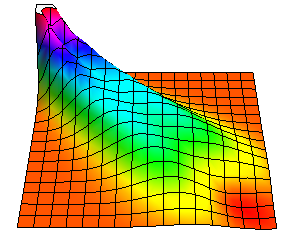}
\caption{$\underline{v}^{20,21}$ Coulomb-Sturmian matrix elements of the Malflet-Tjon potential.}
\label{step1}
\end{figure}

\begin{figure}
\centering
\includegraphics[width=8.5cm]{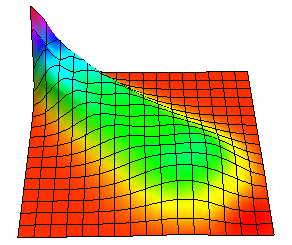}
\caption{$\underline{v}^{20,22}$ Coulomb-Sturmian matrix elements of the Malflet-Tjon potential.}
\label{step2}
\end{figure}

\begin{figure}
\centering
\includegraphics[width=8.5cm]{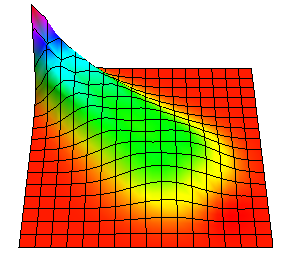}
\caption{$\underline{v}^{20,27}$ Coulomb-Sturmian matrix elements of the Malflet-Tjon potential.}
\label{step7}
\end{figure}

Figures \ref{step1}, \ref{step2} and \ref{step7} display $\underline{v}^{20,21}$, $\underline{v}^{20,22}$ and 
$\underline{v}^{20,27}$, respectively. We can see that as $N'$ increases, the matrix elements around the corner 
become more and more suppressed. Thus if we truncate the matrix to $N\times N$ size, we neglect
terms which are small. We can also see from these pictures that this procedure 
in accordance with
Eq.\ (\ref{gm1}) modifies mostly the lower right corner of the potential matrix. It is also interesting to see 
that in Fig.\ \ref{step1}, even stepping out just by one basis state, and truncating back,  results in a dramatic reduction 
of the matrix elements around the lower right corner. 

\begin{figure}[!ht]
\label{convfig}
\centering
\includegraphics[width=8.5cm]{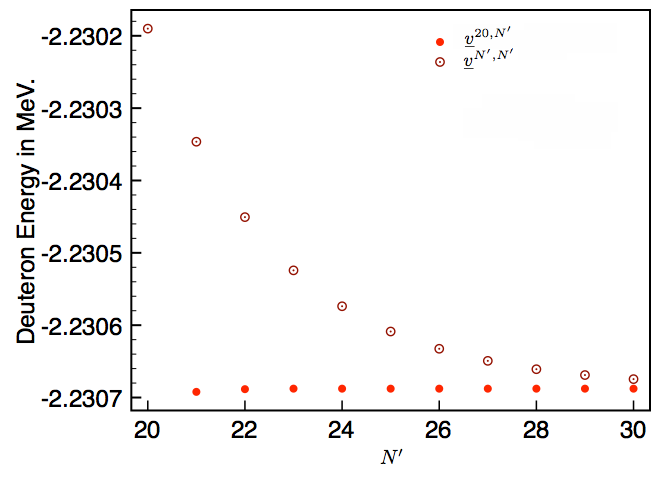}
\caption{Convergence of the deuteron bound state energy with potential matrices 
$\underline{v}^{20,N'}$ and $\underline{v}^{N',N'}$.}
\label{deut}
\end{figure}

Figure \ref{deut} shows the convergence of the deuteron binding energy with 
$\underline{v}^{20,N'}$ and $\underline{v}^{N',N'}$
as a function of $N'$. We can see that our approach of inverting and cutting back the potential matrix
is more advantageous than keeping the original bigger matrix.
We can also see that beyond $N'=N+4 \to N'=N+7$ there is no further improvement. 
We found the same effect with other $N$ values and for scattering states as well. 
So, we fix $N'=N+7$. 

Table \ref{table2} shows the convergence of the deuteron states and $p-p$ scattering phase shifts at 
low, intermediate and high energies with increasing $N$. 
We can observe excellent results even with very low $N$. We observe four digit accuracy with basis representation as low as $N=5$ and eight or nine digit accuracy with $N=20$. The rate of convergence is  better than with the double basis method and higher accuracy can be achieved.

\begin{table}
\caption{The convergence of the deuteron bound state energy and $p-p$ scattering phase shifts at low, intermediate and high energies. The $N\times N$ representation of the potential was calculated from an $N+7 \times N+7$ representation.  }
\label{table2}
\begin{tabular}{| l | | c | c | c | c |  }
\hline
 N & E$_{\mbox{d}}$ & E = 0.1 MeV & E = 1.0 MeV &  E = 100 MeV  \\
\hline
\hline
3&	-2.14459561 &	-0.121277396 &	-0.708187417 &	 	0.378017184 \\
4&	-2.23413092 &	-0.119071149 &		-0.701325274 &	 	0.400336999 \\
5&	-2.22996195 &	-0.119221696 &	-0.701844279 &	 	0.406882711 \\
6&	-2.22826603 &	-0.119221074 &	-0.701711392 &	 	0.406989848 \\
7&	-2.22954511 &	-0.119209853 &	-0.701708254 &	 	0.406858120 \\
8&	-2.23027115 &	-0.119172562 &	-0.701542559 &	 	0.407291677 \\
9&	-2.23060304 &	-0.119165639 &	-0.701519683 &	 	0.407488351 \\
10&	-2.23068178 &	-0.119161793 &	-0.701503458 &	 	0.407494569 \\
11&	-2.23069092 &	-0.119161850 &	-0.701503994 &	 	0.407494183 \\
12&	-2.23068711 &	-0.119161857 &	-0.701503930 &	 	0.407498147 \\
13&	-2.23068566 &	-0.119161942 &	-0.701504323 &	 	0.407498408 \\
14&	-2.23068594 &	-0.119161930 &	-0.701504270 &	 	0.407499317 \\
15&	-2.23068671 &	-0.119161903 &	-0.701504180 &	 	0.407499210 \\
16&	-2.23068728 &	-0.119161894 &	-0.701504142 &	 	0.407499294 \\
17&	-2.23068757 &	-0.119161880 &	-0.701504093 &	 	0.407499379 \\
18&	-2.23068769 &	-0.119161881 &	-0.701504095 &	 	0.407499377 \\
19&	-2.23068773 &	-0.119161878 &	-0.701504085 &		0.407499387 \\
20&	-2.23068774 &	-0.119161880 &	-0.701504090 &	 	0.407499381 \\
21&	-2.23068774 &	-0.119161879 &	-0.701504089 &	 	0.407499381 \\
22&	-2.23068774 &	-0.119161880 &	-0.701504090 &	 	0.407499382 \\
23&	-2.23068774 &	-0.119161880 &	-0.701504090 &	 	0.407499380 \\
24&	-2.23068774 &	-0.119161880 &	-0.701504091&		 	0.407499380 \\
25&	-2.23068774 &	-0.119161880 &	-0.701504091&		 	0.407499380 \\
  \hline
\end{tabular}\label{table:accuracy_check}
\end{table}

\section{Summary and conclusions}
\label{summary}
In this work we propose a new finite-basis representation for the potential operator. 
The approach is inspired by our recent finding concerning Green's operators. If the asymptotic 
Hamiltonian is represented in a discrete basis, then for the resolvent the $\infty \times \infty$ 
symmetric band matrix is inverted by a matrix continued fraction.
A general potential operator is not exactly an infinite band matrix, but it is similar.
The potential matrix exhibits a ridge-like structure which
looks like a band matrix. 

We propose a numerical procedure for a finite-basis representation of the potential such that it retains 
some information about the whole Hilbert space.
We need to calculate the matrix elements of the potential in a
slightly larger basis, about $5-7$ terms larger, invert the matrix numerically, then truncate the matrix to 
the desired size, and finally invert again. This procedure is very straightforward, automatic and 
results in a fast convergence in $N$.

\bibliography{inversion}

\end{document}